\documentclass[aps,twocolumn]{revtex4-1}
\usepackage{graphicx,epsfig}
\usepackage{amsmath}
\usepackage[T1]{fontenc}
\usepackage{hyperref}
\hypersetup{
     colorlinks   = true,
     citecolor    = magenta,
     linkcolor    = blue,
}

\begin{document}

\title{Quantum phases of a spin-1 ultracold Bose gas with three body interactions}
\author{Sk Noor Nabi}%\email{sk.noor@iitg.ernet.in} 
\author{Saurabh Basu} \email{saurabh@iitg.ernet.in} \affiliation{Department of
    Physics, Indian Institute of Technology Guwahati, Guwahati, Assam
    781039, India} \date{\today}

\begin{abstract}
We study the effects of both a repulsive and an attractive three body interaction potential on a spin-1 ultracold Bose gas using mean field approach (MFA). For an antiferromagnetic (AF) interaction, we have found the existence of the odd-even asymmetry in the Mott insulating (MI) lobes in presence of both the repulsive two and three body interactions. In case of a purely three body repulsive interaction, the higher order MI lobes stabilize against the superfluid phase. However, the spin nematic (singlet) formation is restricted upto the first (second) MI lobes for the former one, while there is neither any asymmetry nor spin nematic (singlet) formation is observed for the later case. The results are confirmed after carefully scrutinizing the spin eigen value and spin nematic order parameter for both the cases. On the other hand, for an attractive three body interaction, the third MI lobe is predominantly affected, where it completely engulfs the second and the fourth MI lobes at large values of the interaction strength. Albeit no significant change is observed beyond the fourth MI lobe. In the ferromagnetic case, the phase diagram shows similar features as that of a scalar Bose gas. We have compared our results on the MFA phase diagrams for both types of the interaction potential via a perturbation expansion in both the cases.   
\end{abstract}

\keywords{spinor ultra-cold atoms, three body interaction, magnetic
  field}

\maketitle

\section{Introduction}
\label{intro}
Ultracold atoms trapped in optical lattices offer a scrupulous control over various experimental parameters and hence serve as a supreme arena to explore the quantum many body phenomena. The successful observation of the superfluid (SF) to Mott insulator (MI) transition by Greiner {\it{et al.}} \cite{Greiner} using the spin polarized bosonic $^{87}Rb$ atoms trapped in optical lattices is an elegant observation of the quantum phase transition (QPT). Such efforts were initiated following the proposal by Jacksh {\it{et al.}} \cite{PhysRevLett.81.3108} that the cold atoms in optical lattices can be mapped onto a Bose Hubbard model (BHM), thereby emphasizing how the so-called phase transition can be controlled by tuning the tunneling amplitude and the onsite interparticle interaction strength. 
\\\indent The proliferation in the study of QPT seems to bloom in recent days after the cold atom communities shift their focus to study the quantum degenerate Bose gas with hyperfine spin degrees of freedom. With the technological revolution, a fist step towards this realization is to employ the optical trapping mechanism which helps the neutral atoms to preserve their hyperfine degrees of freedom and hence qualify to treat them as a spinor Bose gas. This is in contrast to a magnetic trapping which freezes the spin degrees of freedom and reduces them to a scalar Bose gas. Fortunately, a proposition to achieve a spinor Bose gas is now a reality after the MIT group has confined all the hyperfine states of $^{23}Na$ condensates using an optical dipole trap which opens up a window of opportunities to explore spin dynamics and quantum magnetism \cite{PhysRevLett.80.2027,PhysRevLett.82.2228}.
\\\indent The pioneering representations of a spin-1 ultracold Bose gas were first portrayed simultaneously by Ho \cite{PhysRevLett.81.742} and Machida \cite{ohmi} by generalizing the Gross-Pitaevskii equation using a low energy Hamiltonian for spinor particles. Contrary to the scalar Bose gas, the spinor Bose gas can be described by a vector order parameter and hence all the hyperfine components transform onto each other through a rotational symmetry in spin space. They also layout a change in the ground state structure of the system and analyzed different spin textures and topological excitations.
\\\indent This predication allows several authors to rigorously go through the rich phase properties possessed by the spin-1 BHM (SBHM) owing to the presence of an additional spin dependent interaction term using diverse techniques, such as mean field approach (MFA) \cite{PhysRevA.91.043620,PhysRevB.69.094410,PhysRevB.77.014503}, quantum Monte carlo (QMC) \cite{PhysRevA.82.063602,PhysRevB.84.064529,PhysRevB.88.104509,PhysRevLett.102.140402} and perturbative expansion \cite{PhysRevA.70.043628,PhysRevLett.94.110403,PhysRevA.87.043624} etc. For an antiferromagnetic (AF) spin dependent interaction, it was found that the Mott insulating phase comprises of a spin singlet phase with even occupation densities and a spin nematic phase with odd occupation densities, where the latter is seen beyond one dimension. Also the phase transition from the spin nematic MI to the SF phase is always second order while spin singlet MI to SF phase corresponds to first order. Besides, the formation of the spin nematic-singlet phase which is responsible for the odd-even asymmetry in the MI lobes pledge richer characteristics of a spinor Bose gas over a scalar gas. \\\indent However the density matrix renormalization group (DMRG) studies done in one dimension outlined that the MI phase with odd occupation densities seem to be a dimerized phase, and there is no signature of a first order transition for the spin singlet MI-SF phase \cite{PhysRevA.68.063602,PhysRevLett.90.250402,EPL.63.505,PhysRevLett.95.240404}. Moreover, it displays odd-even asymmetry at higher values of the spin dependent interaction strength \cite{PhysRevLett.95.240404}. Further, the study on the spinor Bose gas appears to progress steadily in a variety of contexts such as in presence of disorder \cite{PhysRevA.83.013605,NoorJPB,PhysRevB.95.235128}, dipolar interactions \cite{,PhysRevLett.97.020401,PhysRevLett.97.130404}, extended interaction \cite{PhysRevB.92.054506,NOORANDP}, magnetic field \cite{PhysRevA.93.063607,PhysRevA.94.063613,PhysRevLett.84.1066,PhysRevLett.87.080401,Zhang,PhysRevB.70.184434,NoorEPL,PhysRevLett.112.043001}, spin-orbit coupling \cite{PhysRevA.91.023608,PhysRevA.93.013629,PhysRevA.93.023615,PhysRevB.93.081101,PhysRevLett.117.125301} etc. 
\\\indent All of these activities on a spin-1 Bose gas are examined by keeping in mind only the two body spin independent and spin dependent interaction terms which evidently generate curiosity to re-visit their properties in presence of a three and higher body interactions strengths. The effects of repulsive three body interaction on spin-0 (scalar) Bose gas have been subject of intense theoretical study in recent times after Will {\it{et al.}} \cite{Will} claimed to have experimentally observed the multibody interactions \cite{PhysRevA.78.043603,PhysRevA.84.065601,PhysRevA.85.065601,PhysRevA.88.063625}. Unfortunately, effects of such interaction on spin-1 Bose gas are currently lacking and thus motivates us to explore their ground state properties in presence of such three body interaction. 
\\\indent The explicit form of the three body interaction was derived by the Tiesinga {\it{et al.}} \cite{PhysRevA.88.023602} which consists of both the spin independent and dependent terms, similar to that of the two body interaction. Later Silva-Valencia {\it{et al.}} used this model to obtain the phase diagrams of the SBHM with only a three body term \cite{PhysRevA.94.033623} and both the two and three body interactions \cite{arXiv:1707.08195} using a DMRG technique. They claimed that there is no signature of the odd-even asymmetry in the MI lobes in both the cases which sharply contradicts the results for a spinor Bose gas predicted in presence of a two body interaction potential. 
\\\indent This surprising fact gives us an opportunity to study the spin-1 ultracold Bose gas by considering both two as well as three body interaction terms using site decoupling MFA to compare and contrast their claims. Here we found that there exists an odd-even asymmetry in the MI lobes, but the formation of the spin singlet-nematic MI phase is only restricted upto the second MI lobes. 
\\\indent In a rare move, we extend our calculation for a specific type of attractive three body interaction on the SBHM to see the consequences on the phase diagrams, particularly the existence of odd-even asymmetry in the MI lobes. Such a three body interaction term which mostly affects the third MI lobe was earlier studied by Safavi-Naini {\it{et al.}} \cite{PhysRevLett.109.135302} on a scalar Bose gas and proposed a possible way to experimentally observe such type of interactions.
\\\indent This paper is organized as follows. In section II, we
outline our theoretical model for a spinor Bose gas in presence of a three body interaction potential. In section III, we present the phase diagrams of the system for both types of three body interactions using the MFA and perturbative approach. Finally, in section IV we have concluded citing the importance of the results obtained bu us.    
%%%%%%%%%%%%%%%%%%%%%%%%%%%%%%%%%%%%%%%%%%%%
\section{Model}
The Hamiltonian for spin -1 ultracold atoms in presence of both a repulsive and an attractive three body interaction (in addition to the usual two body interaction) can be written as
\cite{PhysRevLett.81.742,ohmi},
\begin{eqnarray}
H&=&-t\sum\limits_{<ij>}\sum\limits_{\sigma}(a^{\dagger}_{i\sigma}a_{j\sigma}+h.c)+\frac{U_{2}}{2}\sum\limits_{i}({\bf{S}}^{2}_{i}-2n_{i})\nonumber \\ &-&\mu
\sum\limits_{i} n_{i}+\frac{U_{0}}{2}\sum\limits_{i}n_{i}(n_{i}-1)+H^{p}_{3}
\label{bhm1}
\end{eqnarray}
where $a^{\dagger}_{i\sigma}$ is the boson creation operator with spin, $\sigma=\pm 1, 0$ and $n_{i}=\sum\nolimits_{\sigma}n_{i\sigma}$, $n_{i\sigma}=a_{i\sigma}^{\dag}a_{i\sigma}$ is the number operator at a site $i$. $t$ is the hopping amplitude from site $i$ to site $j$ and $\mu$ is the chemical potential. The total spin at a site $i$ is given by,
${\bf{S}}_{i}=a^{\dagger}_{i\sigma}{\bf{F}}_{\sigma\sigma'}a_{i\sigma'}$
where ${\bf{F}}_{\sigma\sigma'}$ are the components of spin-1 matrices. $U_{0}$ and $U_{2}$ are the two body spin independent and spin dependent on-site interactions respectively. $H^{p}_{3}$ corresponds to the repulsive \cite{PhysRevA.88.023602} and attractive \cite{PhysRevLett.109.135302} three body interactions respectively which are assumed to have the form,
\begin{equation}
H^{R}_{3}=\frac{W}{6}\sum\limits_{i}n_{i}(n_{i}-1)(n_{i}-2)+\frac{V}{6}\sum\limits_{i}({\bf{S}}^{2}_{i}-2n_{i})(n_{i}-2) 
\end{equation}
\begin{equation}
H^{A}_{3}=U_{3}\sum\limits_{i}\delta_{n_{i},3}
\end{equation}
where the index $p$ refers to the repulsive (R) as well as the attractive (A) three body interaction potentials. $W$ and $V$ are the repulsive three body spin independent and dependent interaction strengths and $U_{3}$ is the attractive three body interaction strength. It was found that the repulsive three body interaction
strength is related to the two body interaction strength by, $W \propto
(V_{0}/E_{r})^{3/4}a^{2}_{s}k^{2}U^{2}_{0}$ ($a_{s}$: $s$ wave scattering length, and $k$: wave vector) \cite{PhysRevA.78.043603} and experimentally $a^{2}_{s}k^{2}$ is
of the order of $10^{-2}$ to $10^{-8}$ \cite{Pethick}. Thus it is reasonable to consider $W<<U_{0}$ and the relationship,
$V/U_{0}=2(U_{2}/U_{0})(W/U_{0})$ only holds for $W<<U_{0}$ and
$V<<U_{2}$ \cite{PhysRevA.88.023602} 
\\\indent To study Eq.(\ref{bhm1}), we shall use the mean field
approximation \cite{PhysRevA.70.043628,PhysRevB.77.014503} to decouple the hopping term as,
\begin{eqnarray}
\hat{a}^{\dagger}_{i\sigma}\hat{a}_{j\sigma} \simeq \langle
\hat{a}^{\dagger}_{i\sigma} \rangle \hat{a}_{j\sigma}
+\hat{a}^{\dagger}_{i\sigma}\langle
\hat{a}_{j\sigma}\rangle-\langle
\hat{a}^{\dagger}_{i\sigma}\rangle \langle
\hat{a}_{j\sigma}\rangle
\label{deco}
\end{eqnarray}
Now defining the superfluid order parameter as the equilibrium value of an
operator at a site $i$ as $\psi_{i\sigma}= \langle \hat{a}_{i\sigma}\rangle$, then Eq.(\ref{bhm1}) can be written as the sum of the mean field Hamiltonians, $H=\sum\nolimits_{i}H^{MF}_{i}$ where $H^{MF}_{i}$ is given by,
\begin{eqnarray}
H^{MF}_{i}&=&\underbrace{\frac{U_{0}}{2}\hat{n}_{i}(\hat{n}_{i}-1)+\frac{U_{2}}{2}({\bf{S}}^{2}_{i}-2\hat{n}_{i})-\mu\hat{n}_{i}+H^{p}_{3}}_\text{$H^{0}$}\nonumber \\
&-&\underbrace{t\sum\limits_{\sigma}(\psi_{i\sigma}\hat{a}_{i\sigma}+h.c)+t\sum\limits_{\sigma}\psi_{i\sigma}^{2}}_\text{$H^{'}$}
\label{mf}
\end{eqnarray}
In order to study the SF-MI phase transition, we need to find the equilibrium values of the SF order parameter, $\psi^{eq}_{i}$ and local densities, $\rho^{eq}_{i}$. To accomplish the task, we first form the matrix elements of the mean field Hamiltonian, $H^{MF}_{i}$ in the site occupation number basis, $|\hat{n}_{i\sigma}\rangle$ as $\langle\hat{n}_{i\sigma}|H^{MF}_{i}|\hat{n}'_{i\sigma}\rangle$, hence diagonalize it starting with some guess values for $\psi_{i\sigma}$ with $n_{i}=7$ and continued this process until the self consistency condition for $\psi^{eq}_{i}$ is reached. Finally, we compute the equilibrium SF order parameter and local densities as, 
\begin{equation} 
\psi_{i\sigma}=\langle\Psi_{g}|a_{i\sigma}|\Psi_{g}\rangle;\hspace{1cm} 
\rho_{i\sigma}=\langle\Psi_{g}|n_{i\sigma}|\Psi_{g}\rangle
\end{equation}
where we will drop {\it{eq}} from here and the total SF order parameter is, $\psi_{i}=\sqrt{\psi^{2}_{i+}+\psi^{2}_{i0}+\psi^{2}_{i-}}$.
%%%%%%%%%%%%%%%%%%%%%%%%%%%%%%%%%%%%%%%%%%%%
\section{Results}
\indent In order to see the effect of both the repulsive and attractive three body
interaction on a spin-1 Bose gas, we first consider the atomic limit, that is, $t=0$ in Eq.(\ref{bhm1}). In
the atomic limit, the system is completely in the insulating phase with an
energy gap, $E_{g}$ in the particle-hole excitation spectra, which is
the difference between the upper ($\mu_{+}$) and the lower ($\mu_{-}$)
values of the chemical potential corresponding to a MI lobe for a given occupancy, $n$ \cite{PhysRevA.83.013605}. For $t=0$, corresponding to the repulsive three body interaction case, the energy eigen values, $E^{0}(S,n)$ of $H^{0}$ consists of only the unperturbed terms, that is,
\begin{eqnarray}
E^{0}(S,n)&=&\frac{W}{6}n(n-1)(n-2)+\frac{V}{6}[S(S+1)-2n](n-2)\nonumber\\
&-&\mu n+\frac{U_{0}}{2}n(n-1)+\frac{U_{2}}{2}[S(S+1)-2n]
\end{eqnarray} 
\begin{figure}[!t]
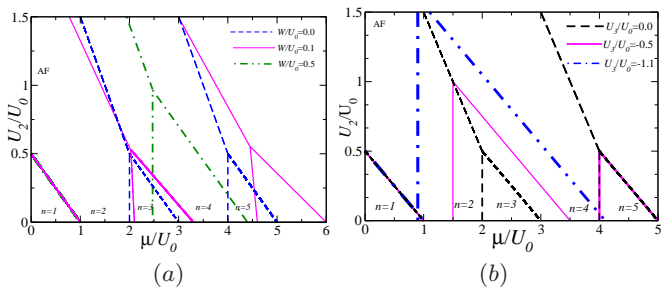

  \centerline{ \hfill
    \psfig{file=1a.eps,width=0.24\textwidth}
    \hfill \hfill \hspace{0mm}\psfig
           {file=1b.eps,width=0.24\textwidth}
           \hfill} \centerline{\hfill $(a)$ \hfill\hfill $(b)$ \hfill}
  \caption{In the atomic limit ($t=0$), the MI lobe widths for three body repulsive in (a) and attractive interaction potentials in (b). The odd-even asymmetry exists in Fig.(a) while in Fig.(b), no asymmetry is observed. }
\label{1}
\end{figure}
which has a common eigenstate $|S,S_{z},n\rangle$ where the
corresponding operators, namely $S,S_{z},n$ commute with each other.  
Assuming the spin eigen value, $S=0$ for the even and $S=1$ for the odd MI lobes, similar to that in Ref.\cite{PhysRevA.83.013605}, the boundary of the MI lobe ($\mu_{\pm}$)
can be found from the relation,
$E^{0}(S_{1},n_{1})<E^{0}(S,n)<E^{0}(S_{2},n_{2})$, where $S_{1,2},n_{1,2}$ 
are the lower and higher spin and density values respectively corresponding to particular $S,n$
values. In the AF case, this inequality gives the following conditions, which can be stated as, 
\\(i) For the odd MI lobes ( that is, $n=1,3,...$):
$(n-1)+(n-1)(n-2)W/2U_{0}+(1-n)V/3U_{0}<\mu/U_{0}<n-2U_{2}/U_{0}+n(n-1)W/2U_{0}-(n-1)V/U_{0}$. If
we equate these two $\mu$ values, we shall obtain a critical
$U_{2}/U_{0}$, given by $U^{c}_{2}/U_{0}=1/2+(n-1)[W/2U_{0}-V/3U_{0}]$ below
which the odd MI lobes exist and above which the odd MI lobes vanish.
\\(ii) For even MI lobes ( that is, $n=2,4,..$): If
$U_{2}/U_{0}<U^{c}_{2}/U_{0}$, then
$(n-1)-2U_{2}/U_{0}+(n-1)(n-2)W/2U_{0}+(2-n)V/U_{0}<\mu/U_{0}<n+n(n-1)W/2U_{0}-nV/3U_{0}$.
For $U_{2}/U_{0}>U^{c}_{2}/U_{0}$,
$n-3/2-U_{2}/U_{0}+(n-2)^{2}W/2U_{0}+2(2-n)V/3U_{0}<\mu/U_{0}<n+1/2-U_{2}/U_{0}+n^{2}W/2U_{0}-2nV/3U_{0}$.
\\ \indent If we plot all these equations for different values of
$W/U_{0},V/U_{0}$, we shall obtain the structures for the MI lobes as shown in
Fig.\ref{1}(a). At $W/U_{0}=0.1$, which yields
$V/U_{0}=0.2U_{2}/U_{0}$, the even MI lobes become
more stable compared to the odd MI lobes, and the width of the chemical
potential, $\mu$ for all the MI lobes, except the first one, increases
with the three body interaction strength, $W$. Interestingly, the critical $U^{c}_{2}/U_{0}$ for the disappearance of all odd MI lobes in the absence of $W/U_{0}$ was 0.5
\cite{PhysRevA.83.013605}, which now changes to 0.53 and 0.553 at
$W/U_{0}=0.1$ corresponding to the third and the fifth odd MI lobes
respectively. We have also considered a higher value of $W/U_{0}$, namely, $W/U_{0}=0.5$, 
for which $V/U_{0}=0.05\sim U_{2}/U_{0}$ and found that the chemical
potential widths get enhanced, thereby making inroads for the MI
phase. Thus the MI lobes become more stable compared to the SF phase. This clearly suggests 
of an existence of an odd-even asymmetry in the successive MI lobes even in presence of the three body interaction. We shall discuss this in a shortwhile. 
\begin{figure}[!t]
  \centerline{ \hfill
\psfig{file=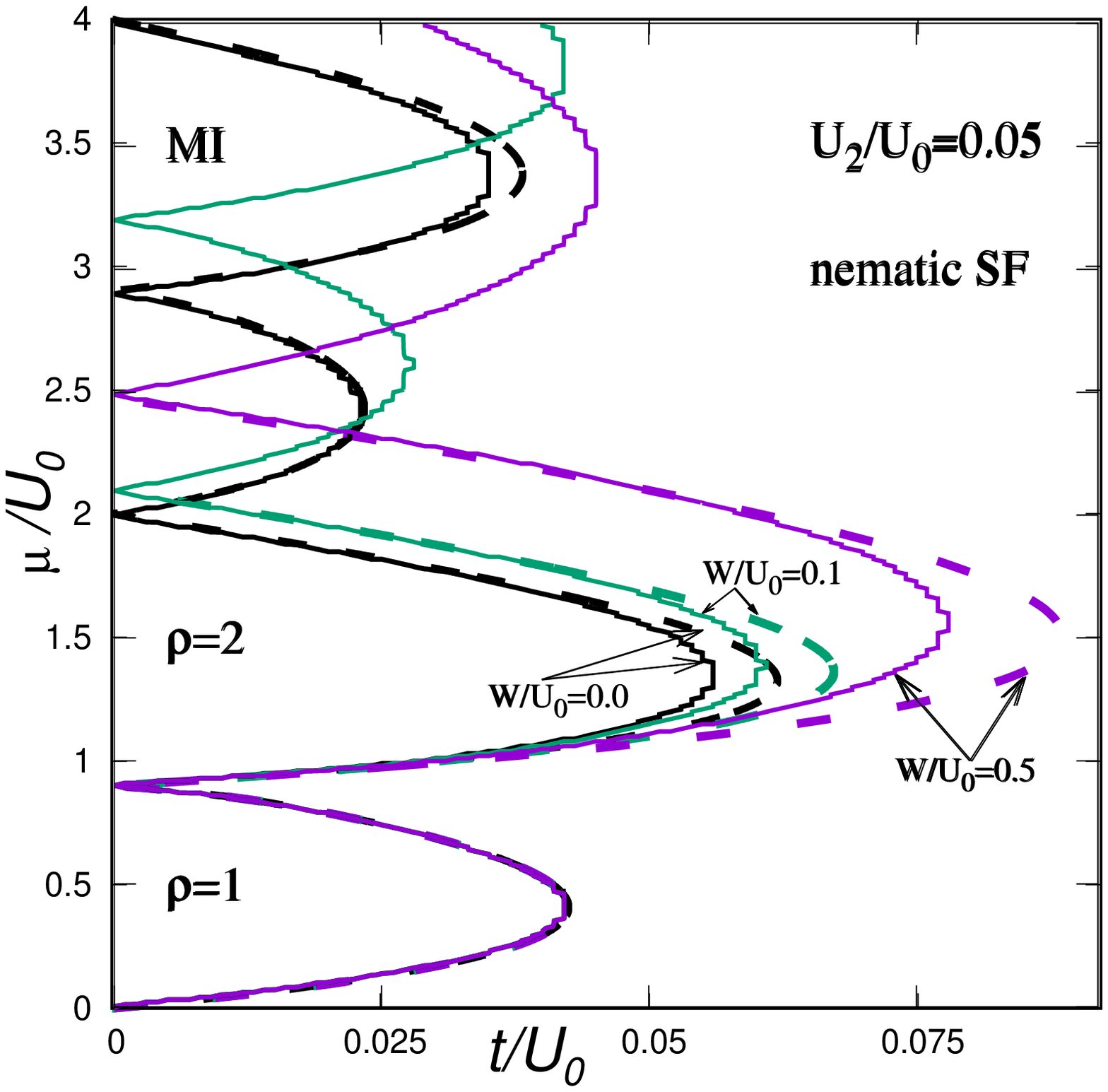,width=0.24\textwidth}
    \hfill\hfill 
      \psfig{file=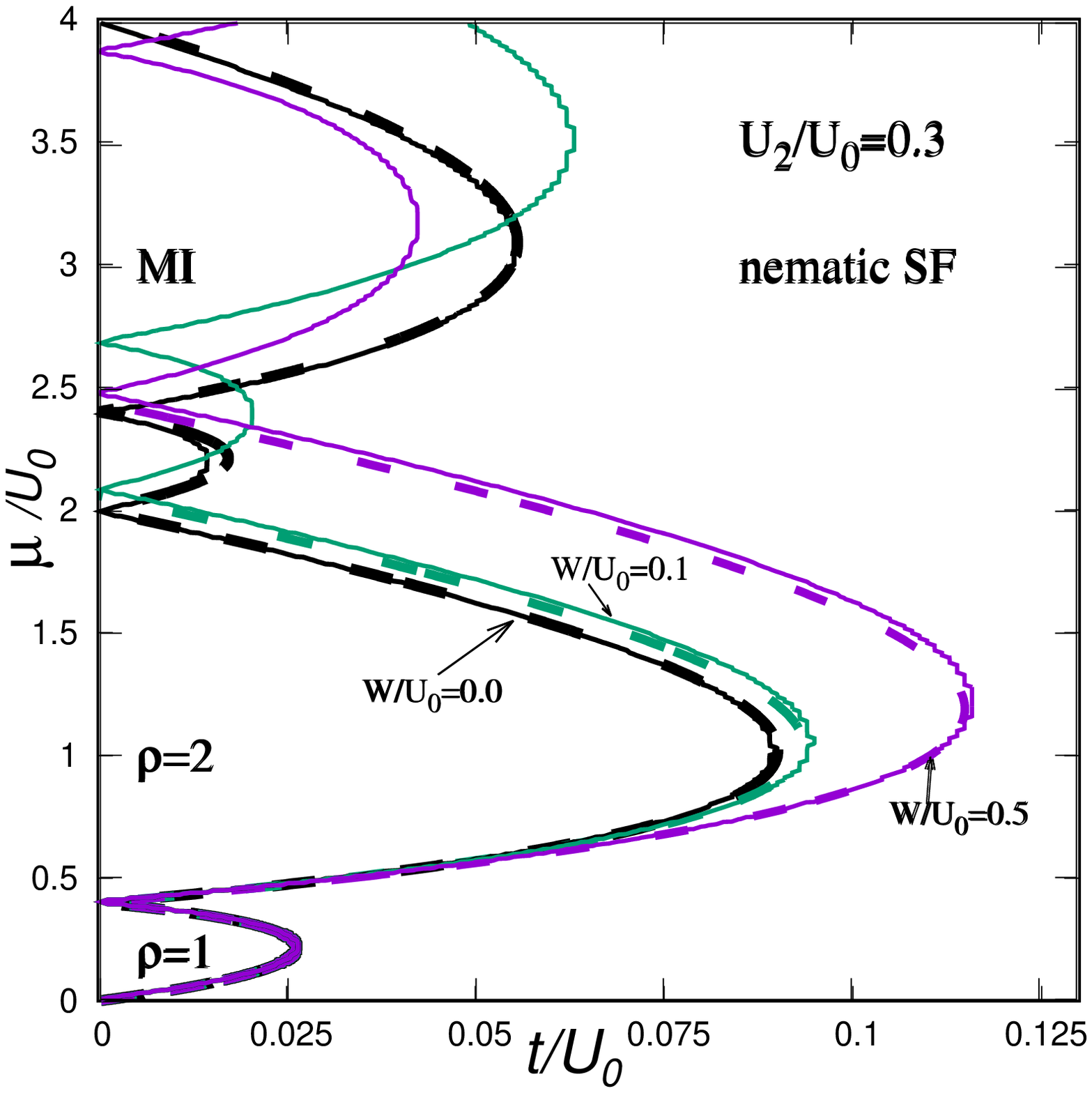,width=0.235\textwidth}    
         \hfill} 
\centerline{\hfill $(a)$ \hfill\hfill $(b)$
    \hfill}
\centerline{ \hfill
\psfig{file=2cd.eps,width=0.485\textwidth}           
         \hfill} 
\centerline{\hfill $(c)$ \hfill\hfill $(d)$ \hfill}
  \caption{Phase diagrams with $W/U_{0}$ for $U_{2}/U_{0}=0.05$ in (a) and $U_{2}/U_{0}=0.3$ in (b). In both cases, such asymmetry is observed. The population fraction, $N_{\sigma}/N$ with occupation densities, $\rho$ is shown in (c) which indicates that the spin singlet formation is only possible at the second MI lobe. }
\label{2}
\end{figure}  
\\\indent Now we shall include the attractive three body interaction, 
$H^{A}_{3}$ in $H^{0}$ to see how it effects the MI phase. Again assuming 
same spin eigen values for the odd and even MI lobes, we have obtained the 
boundaries of each of the MI lobes individually and plotted them in Fig.\ref{1}(b) for 
different values of $U_{3}/U_{0}$ corresponding to the AF case. It shows that at a representative value for $U_{3}$, namely, $U_{3}/U_{0}
=-0.5$, although there is no change in the first and the fifth MI lobes but 
surprisingly the third MI lobe expands considerably, thereby engulfing the 
second and the fourth MI lobes. This result is in contrast with that for the repulsive
three body interaction. Also, while the critical values for $U_{2}/U_{0}$ for the 
first and the fifth MI lobes remain at 0.5 however the third MI lobe vanishes 
when $U^{c}_{2}>U_{0}/2-U_{3}$. Further, for a larger value of 
$U_{3}/U_{0}=-1.1$, the third MI lobe grows larger compared to the other 
MI lobes by completely encroaching into the second and the fourth MI lobes. This implies 
that there exists a critical value of $U_{3}$ below which the second and the fourth MI lobes survive, and it corresponds to $U^{c}_{3}/U_{0}=1+2U_{2}/U_{0}$.      
\\ \indent Similarly for the ferromagnetic case, the MI lobes are identical with a spin-0 (scalar) Bose gas in presence of both the repulsive and the attractive three body interaction potentials. The chemical potential width increases for the second and higher MI lobes with the inclusion of $W/U_{0}$, while the third MI lobe gets affected by the inclusion of the attractive three body interaction term.
\\ \indent Next we turn on the hopping
strength and obtain the phase diagram via the mean field
approximation (MFA) (see Eq.(\ref{mf})) for a complete visualization 
about the existence of the odd-even asymmetry and hence to explore the nature of the MI-SF phase transition in presence of three body interactions. 
\\\indent Recently, the phase diagrams of SBHM in presence of both the two and three body repulsive interaction were obtained by Silva-Valencia {\it{et al.}} \cite{arXiv:1707.08195} using DMRG technique in the AF case, where they have claimed that the 
odd-even asymmetry only exists for small values of the hopping strength, that is, $t/U_{0}$. But as the hopping strength is increased, such asymmetry vanishes and there is a signature of a phase transition from a longitudinal polar (LP) to a transverse polar (TP) SF phase. 
\\\indent To crosscheck their claims, here we will use same parameter values, that is, $U_{2}/U_{0}=0.05$ for the AF case. The MFA phase diagrams are shown in 
Fig.\ref{2}(a) for different values of $W/U_{0}$. At $W/U_{0}=0.1$, we found that although there is no modification of the first MI lobe, the second and the higher MI lobes get enhanced with
$W/U_{0}$ as seen from Fig.\ref{1}(a). On increasing $W/U_{0}$, the MI phase encroaches more into the SF regime, pushing the system towards an insulating phase which results in increase of the location
for the MI-SF phase transition, $t_{c}/U_{0}$. \\\indent We have also considered a higher value for the spin dependent interaction, that is, $U_{2}/U_{0}=0.3$ and the resulting phase diagrams with similar $W/U_{0}$ are shown in Fig.\ref{2}(b). At higher value of $W/U_{0}=0.5$, we have chosen $V=0.1\ne 2U_{2}/U_{0}W$ so that $V<<U_{2}$. This result convincingly demonstrates the existence of the odd-even asymmetry in the insulating phase. The phase diagrams corresponding to $W/U_{0}=0$ included for comparison, were studied earlier in
Refs.\cite{PhysRevA.70.043628,PhysRevB.77.014503}. So it is quite reasonable to 
conclude that the odd-even asymmetry in the MI lobes that exists in the MFA phase diagrams 
with the three body interaction is not an artifact of this method, but rather
emphasizes far rich phase properties that a spinor Bose gas exhibit. 
\\\indent Also to verify the signature of the LP to TP SF phase transition reported in Ref.\cite{arXiv:1707.08195}, we have plotted the population fraction, $N_{\sigma}/N$ as a function of the occupation densities, $\rho$ for $W/U_{0}=0.1,0.5$ in Fig.\ref{2}(c). At $U_{2}/U_{0}=0.05$, it shows that two hyperfine populations are equal, that is, $N_{+}/N=N_{-}/N$ and $N_{\pm}/N$ dominates over $N_{0}/N$ for all values of $W/U_{0}$. This suggests that the SF phase obtained from the MFA is completely in the transverse polar phase since $N_{\pm}>N_{0}$, and that there is no crossover between the $N_{\pm}/N$ and $N_{0}/N$ indicates an absence of such phase transition. This result is again inconsistent with the finding of Ref.\cite{arXiv:1707.08195}. However, surprisingly, it tells us that the spin singlet formation is only restricted at the second MI lobe ($\rho=2$) where $N_{\pm}/N=N_{0}/N$, while for the fourth MI lobe ($\rho=4)$, the hyperfine fractions differ from each other \cite{PhysRevLett.102.140402}. This issue will be made more clear in the upcoming discussion.
\begin{figure}[!h]
  \centerline{ \hfill
\psfig{file=3a.eps,width=0.255\textwidth}
    \hfill\hfill 
      \psfig{file=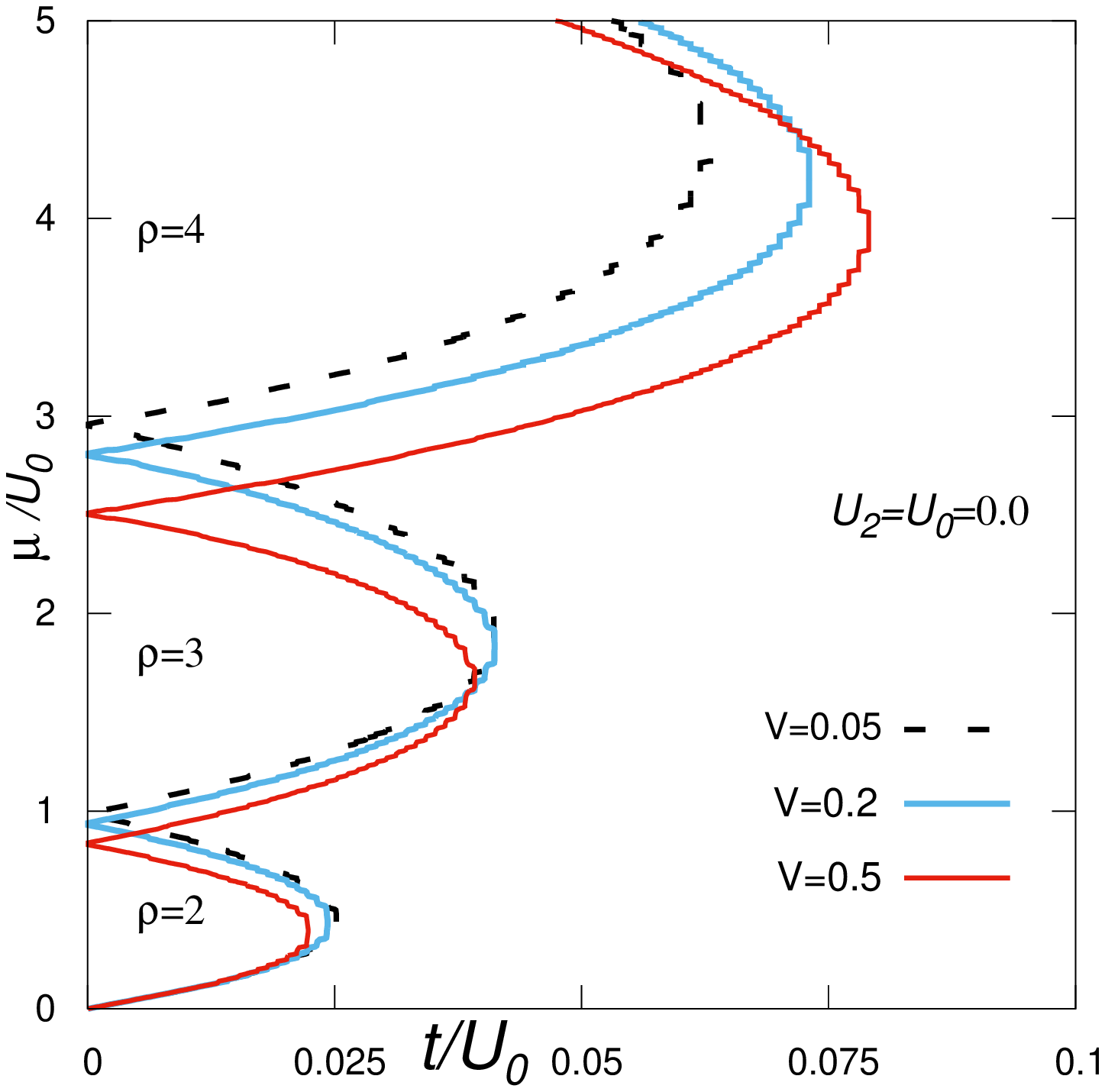,width=0.22\textwidth}    
         \hfill} 
\centerline{\hfill $(a)$ \hfill\hfill $(b)$
    \hfill}
  \caption{1D variation of $\rho$ with $W/U_{0}$ for $U_{2}/U_{0}=0.05$ and $U_{2}/U_{0}=0.56$ in (a). Phase diagram with purely three body repulsive interaction strength, $V$ with $W=1$ and $U_{2}=U_{0}=0.0$ in (b).}
\label{3}
\end{figure}
\\\indent In Fig.\ref{3}(a), we have shown variations of the local density, $\rho$ as a function of the chemical potential to ascertain the order of MI-SF phase transition for different values of $W/U_{0}$. We have found that for smaller values of $U_{2}/U_{0}=0.05$, the MI-SF phase transition still maintains a first order character for the even, and a second order for the odd MI lobes at both the values of $W/U_{0}$ as pointed out earlier in Refs.\cite{PhysRevB.77.014503,PhysRevB.88.104509} which do not include a three body interaction term. For larger $U_{2}/U_{0}=0.3$, there is no signature of the first order transition for the even MI lobes, as it was found in Ref.\cite{PhysRevB.88.104509}, but all the MI-SF phase transitions are of second order in nature. We have also checked that the first and third odd MI lobes disappear when the spin dependent interaction value satisfies, $U_{2}/U_{0}=0.53\ge U^{c}_{2}/U_{0}$ at $W/U_{0}=0.1$ and $U^{c}_{2}/U_{0}=0.967$ for $W/U_{0}=0.5$ [Fig.\ref{3}(a)]. 
\\\indent It is also quite interesting to investigate the effects of only the repulsive three body interaction (without a two body interaction) on a spin-1 ultracold Bose gas, which was studied in Ref.\cite{PhysRevA.94.033623} using DMRG technique. For this purpose, we set $U_{2}=U_{0}=0$, $W=1$ and the resulting MFA phase diagrams with different values of the three body spin dependent interaction, $V$ are shown in Fig.\ref{3}(a). The results show that although the MI lobe with unit occupancy ($\rho=1$) is absent, however the second and the higher order MI lobes are more stable than the SF phase. At larger values of the spin dependent three body term, $V$, that is at $V=0.5$, the chemical potential widths of the second and the third MI lobes shrink, thereby resulting in the stability of the fourth MI lobes. This clearly indicates that the odd-even asymmetry is completely absent when only the three body interaction is present and these phase diagrams are in qualitative agreement with the results in Ref.\cite{PhysRevA.94.033623}. Finally, by comparing the results of Fig.\ref{2}(a)(b) and Fig.\ref{3}(a), we conclude that the two body spin dependent interaction is solely responsible for the odd-even asymmetry in the MI lobes, while the three body interaction just adds to the stability of the insulating phases.
\begin{figure}[!h]
  \centerline{ \hfill
    \psfig{file=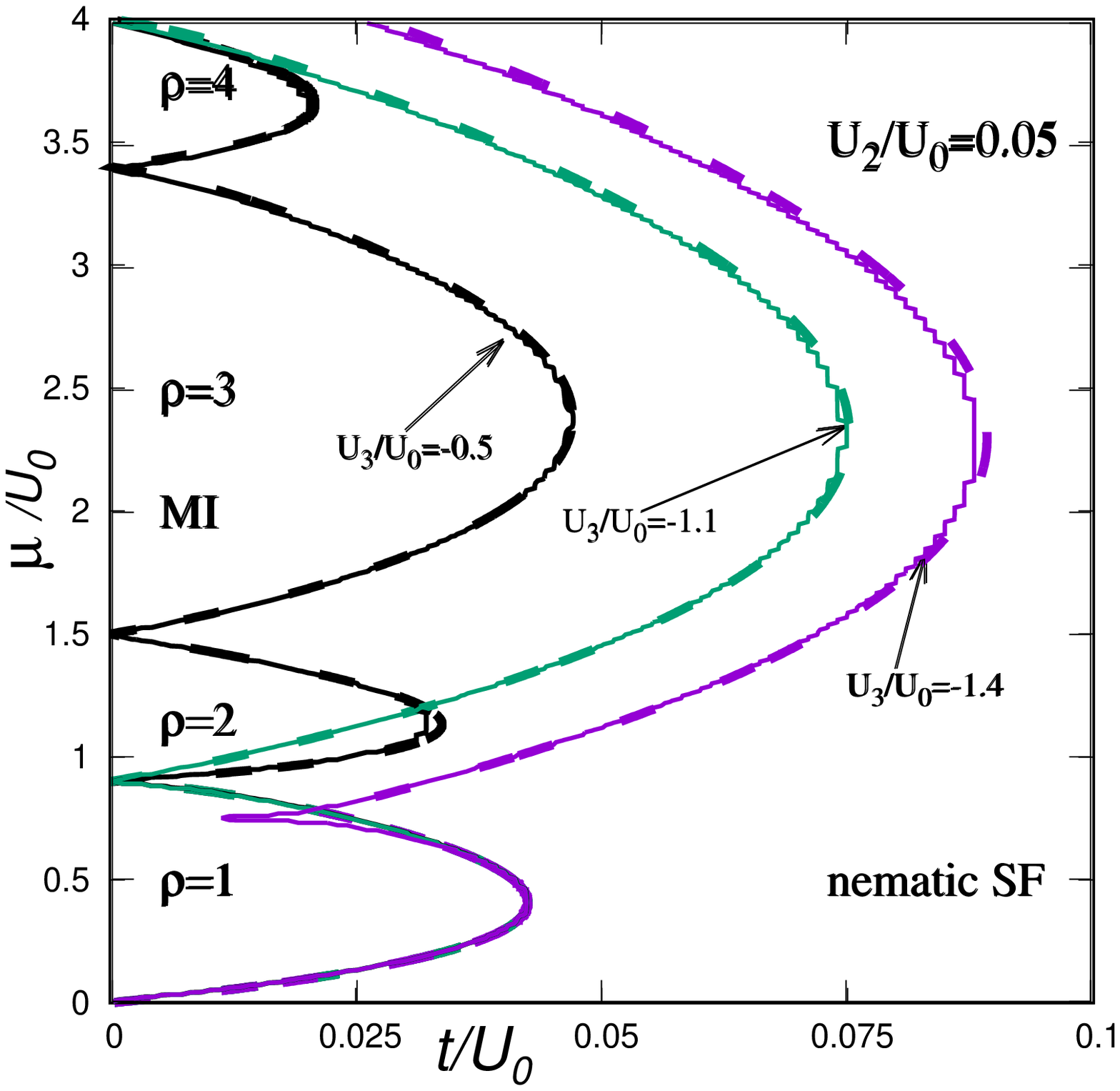,width=0.24\textwidth}
    \hfill \hfill \hspace{0mm}\psfig
           {file=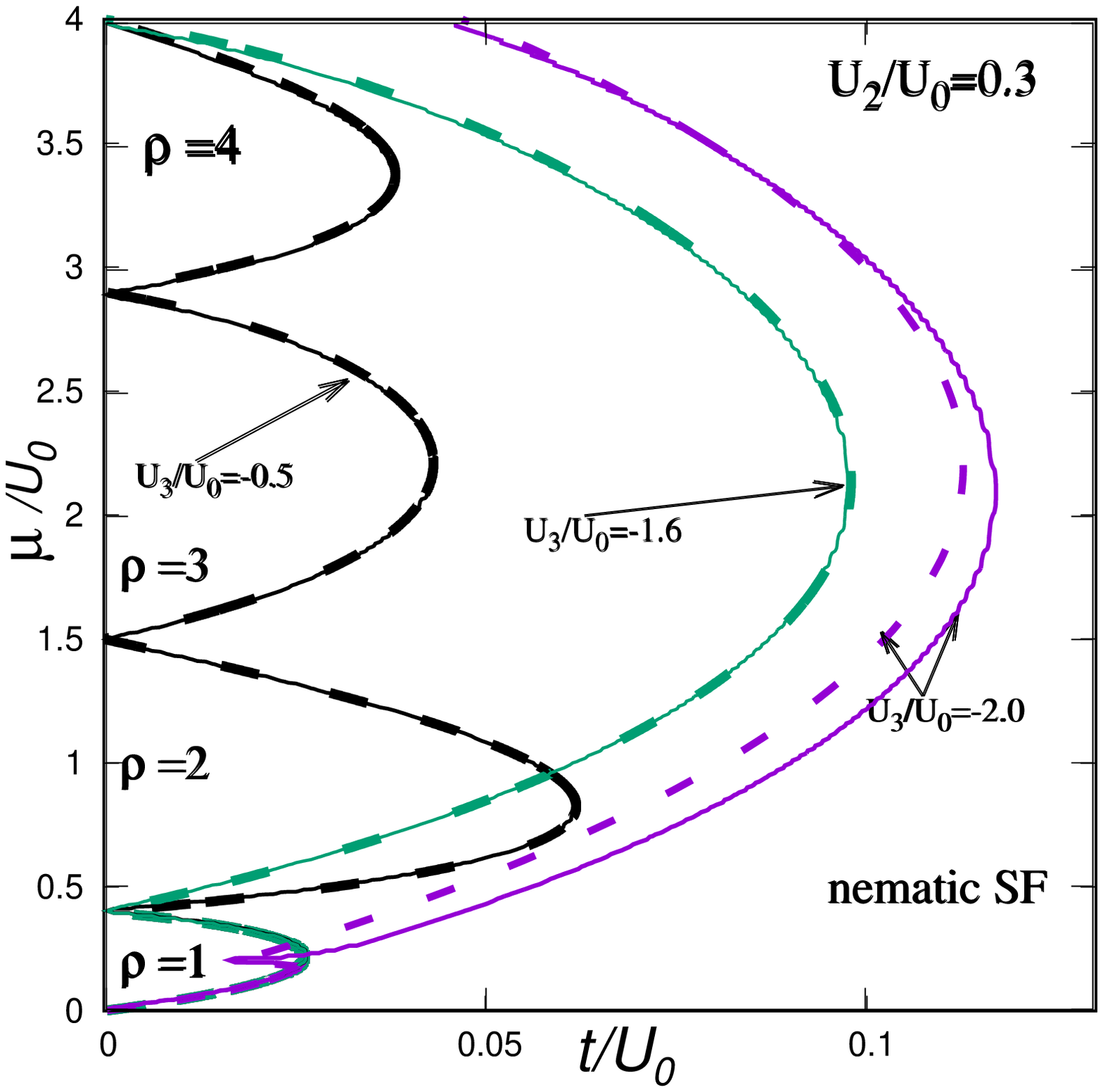,width=0.24\textwidth}
           \hfill} \centerline{\hfill $(a)$ \hfill\hfill $(b)$ \hfill}
  \caption{Phase diagrams in case of an attractive three body interaction in the AF case for $U_{2}/U_{0}=0.05$ in (a) and $U_{2}/U_{0}=0.3$ in (b). In both the cases, the third MI lobe predominantly stabilizes by encroaching into the second and the fourth MI lobes.}
\label{4}
\end{figure}
\\\indent Now we shall consider the attractive three body interaction, $U_{3}/U_{0}$ and the phase diagrams for the AF case are shown in Fig.\ref{4}. For $U_{2}/U_{0}=0.05$ [Fig.\ref{4}(a)], we have found that at $U_{3}/U_{0}=-0.5$, the third MI lobe grows considerably by 
spanning into both the second and the fourth MI lobes, while the first MI lobe remains 
unaffected as seen from Fig.\ref{1}(b). At $U_{3}/U_{0}=-1.1$, the third MI lobe 
completely occupies the second and the fourth MI lobes since the critical value of 
$U_{3}/U_{0}$ below which they exist corresponds to $1+2U_{2}/U_{0}$ as seen from 
Fig.\ref{1}(b). On increasing $U_{3}/U_{0}$, it further stabilizes the third MI 
phase and hence the critical tunneling strength, $t_{c}/U_{0}$ also increases for 
the MI-SF phase transition. Similar effects are observed for higher values of $U_{2}/U_{0}=0.3$ in Fig.\ref{4}(b) corresponding to different values of $U_{3}/U_{0}$.
\begin{figure}[!h]
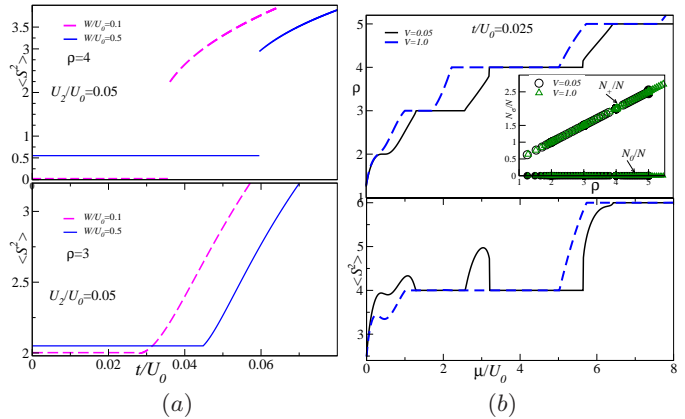

  \centerline{ \hfill
    \psfig{file=5a.eps,width=0.245\textwidth}
    \hfill \hfill \hspace{0mm}\psfig
           {file=5b.eps,width=0.245\textwidth}
           \hfill} \centerline{\hfill $(a)$ \hfill\hfill $(b)$ \hfill}
  \caption{The total spin eigen value, $\langle S^{2} \rangle$ variation in the AF case with
    $W/U_{0}$ corresponding to MI lobe with $\rho=3$ and $\rho=4$ in (a). It shows that the spin singlet formation is impossible for $\rho=4$ MI lobe. The occupation densities, $\rho$ and $\langle S^{2} \rangle$ variation in case of a purely three body repulsive interaction show $\langle S^{2} \rangle=4$ for the $\rho=3$ and $\rho=4$ MI lobes (b).}
\label{5}
\end{figure}
\\\indent In order to support our MFA phase diagrams more convincingly, we now shift our attention to the perturbative mean field approaches (PMFA) for both types of the three body interaction. For this, we first calculate the total spin eigen value, $\langle S^{2} \rangle$ corresponding to the even and odd MI lobes and the results for the AF case are shown in Fig.\ref{5}. 
\\\indent For $U_{2}/U_{0}=0.05$, we have found that $\langle S^{2} \rangle=2.000$ for the first and $\langle S^{2} \rangle=0$ for the second MI lobes, while for the $\rho=3$ lobe, $\langle S^{2} \rangle \simeq 2.0025$ and for the $\rho=4$ MI lobe, $\langle S^{2} \rangle \simeq 0.0282$ at $W/U_{0}=0.1$. On increasing $W/U_{0}=0.5$, $\langle S^{2} \rangle$ remains same for the first and the second MI lobes but has a value $\langle S^{2} \rangle \simeq 2.05$ for the third and $\langle S^{2} \rangle \simeq 0.553$ for the fourth MI lobes. The second and the fourth MI lobes show a first order transition to the SF phase 
[Fig.\ref{5}(a)]. This re-emphasizes our previous prediction of possible spin singlet formation only for the second but not for the fourth MI lobe. This was also outlined earlier from $N_{\sigma}/N$ behavior as a function of $\rho$. At $U_{2}/U_{0}=0.3$, we have observed similar $\langle S^{2} \rangle$ variation for the third and the fourth MI lobes all values of $W/U_{0}$. 
\\\indent To make our point regarding the spin singlet formation more emphatic, we have computed the $z$-component of the spin nematic order parameter, is defined as, $Q_{zz}=
\langle {\bf{S}}^{2}_{z} \rangle-(1/3)\langle {\bf{S}}^{2} \rangle$ \cite{PhysRevLett.88.163001,PhysRevB.70.184434,PhysRevB.88.104509,PhysRevA.68.063602} which is shown in Fig.\ref{6} for different values of $W/U_{0}$ corresponding to all the MI lobes. For $U_{2}/U_{0}=0.05$, it shows that $Q_{zz}$ is finite for the first as well as the third MI lobes and hence shows a second order transition to the SF phase. However for the second MI lobe ($\rho=2$), $Q_{zz}$ completely vanishes in the MI phase and then shows a jump to the SF phase at both values of $W/U_{0}$. While for the fourth ($\rho=4$) MI lobe, we observe that the spin nematic order parameter is finite in the MI phase, but still maintains a first order transition to the SF phase regardless of the value of $W/U_{0}$. At $U_{2}/U_{0}=0.3$, for the second and the fourth MI lobes, $Q_{zz}$ exhibits similar variation in the MI phase and hence registers a second order transition to the SF phase. This finally supports the claim made earlier about the possible spin singlet formation only in the second MI lobe which does not occur in the fourth MI lobe.
\begin{figure}[!t]
  \centerline{ \hfill
    \psfig{file=6a.eps,width=0.49\textwidth}
    \hfill} 
\end{figure}
%\vspace{-1cm}
\begin{figure}[!t]
  \centerline{ \hfill
    \psfig{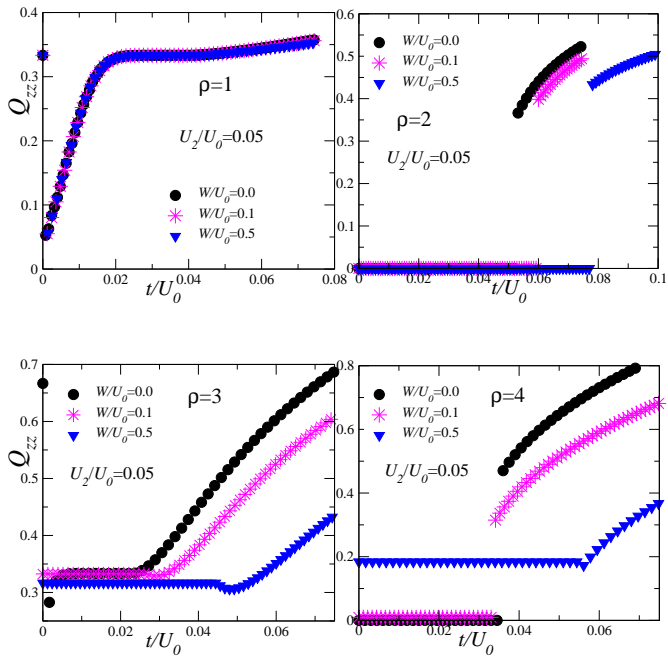}
    \hfill } 
\caption{The spin nematic order parameter, $Q_{zz}$ for all four MI lobes for $U_{2}/U_{0}=0.05$ with different $W/U_{0}$. For the odd MI lobes, $Q_{zz}$ is finite in the MI phase, then shows a second order transition to the SF phase. But in the MI phase for the second lobe $\rho=2$, $Q_{zz}=0$, while it is $Q_{zz}\ne 0$ for the fourth MI lobe ($\rho=4$). Further, $Q_{zz}$ shows a first order transition to the SF phase.}
\label{6}
\end{figure}     
\\\indent But in case of a purely repulsive three body interaction ($U_{2}=U_{0}=0.0$), we have noticed that the $\langle S^{2} \rangle=4$ for the third and the fourth MI lobes, while $\rho=2$ shows an oscillatory behavior of $\langle S^{2} \rangle$ as a function of $V$ as shown in Fig.\ref{5}(b). Further, the population fraction shows that the SF phase is completely in the TP phase since $N_{\pm}/N>N_{0}/N$ [Fig.\ref{5}(b)].
\\ \indent At this point, we are ready to perform the perturbation expansion (PMFA) using the above information on the spin eigen values, $\langle S^{2} \rangle$ to determine the modified ground state energy, $E_{g}$ with $H^{'}$ as the perturbation term [see Eq.(\ref{mf})]. In the AF case, using the same eigenstate for $H^{0}$ from Ref.\cite{PhysRevA.70.043628}, $E_{g}$, after incorporating the first and the second order corrections, can be expressed in a series expansion of $\psi$ of the form,
\begin{eqnarray}
E_{g}(\psi)&=&E^{0}+E^{1}+E^{2}\nonumber\\
&=&E^{0}+C_{2}(U_{0},U_{2},\mu,n,W,V,U_{3})\sum\limits_{\sigma}\psi^{2}_{\sigma}
\end{eqnarray}
Minimization of $E_{g}(\psi)$ with respect to $\psi$ leads to $C_{2}(U_{0},U_{2},\mu,n,W,V,U_{3})=0$ which yields the boundary between the SF and the MI phases.  
\\\indent In presence of two and three body repulsive interaction potentials, the above condition for the even MI lobes with $\langle S^{2} \rangle=0$ yields,
\begin{equation}
\Big{(}\frac{1}{t}\Big{)}=\frac{n/3}{\alpha+2U_{2}-(n-2)[\delta+V]}
+ \frac{(n+3)/3}{\beta+n\delta-nV/3}
\label{even}
\end{equation}
while for the odd MI lobes with $\langle S^{2} \rangle=2$, it corresponds to,
\begin{eqnarray}
\Big{(}\frac{1}{t}\Big{)}&=&\frac{(n+2)/3}{\alpha-(n-2)\delta+\gamma/3}+\frac{(n+1)/3}{\beta-2U_{2}+n\delta-\gamma}
\nonumber
\\ &+&\frac{4(n-1)/15}{\alpha+3U_{2}-(n-2)\delta-(4n-10)V/3}\nonumber
\\ &+&\frac{4(n+4)/15}{\beta+U_{2}+n\delta+2n\gamma/3}
\label{odd}
\end{eqnarray}
where $\alpha=\mu-(n-1)U_{0}$, $\beta=-\mu+nU_{0}$, $\delta=(n-1)W/2$ and $\gamma=(n-1)V$.
\\\indent We have seen that $\langle S^{2} \rangle=2$ only for the first and $\langle S^{2} \rangle=0$ for the second MI lobes and thus we shall restrict our calculation upto the second MI lobe and the resulting phase boundaries are plotted with $W/U_{0}$ in Fig.\ref{2} (dashed lines). At both values of $U_{2}/U_{0}$ and $W/U_{0}$, the analytical phase diagrams are in excellent agreement with the MFA results except at the tip of the second MI lobe. 
\begin{figure}[!h]
  \centerline{ \hfill
    \psfig{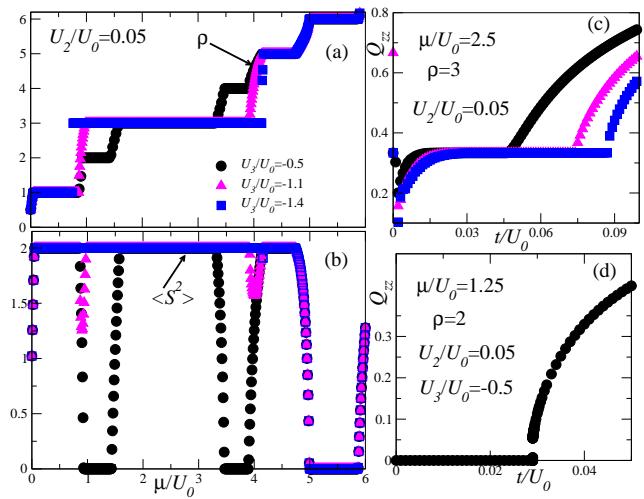}
    \hfill } 
\caption{The occupation density, $\rho$ in (a) and spin eigen value, $\langle S^{2} \rangle$  in (b) for $U_{2}/U_{0}=0.05$ with different values of $U_{3}/U_{0}$. The occupation densities show that odd-even asymmetry remains intact beyond $\rho=4$ MI lobe. The spin nematic order parameter, $Q_{zz}$ for the MI lobes with $\rho=3$ in (c) and $\rho=2$ in (d) with $U_{3}/U_{0}$ shows that the SF phase is a nematic phase.}
\label{7}
\end{figure}    
\\\indent In Fig.\ref{7}(b), we have shown the $\langle S^{2} \rangle$ variation as a function of $\rho$, hence carry out similar PMFA and finally map them with mean field phase diagrams for the attractive three body interaction potential. For $U_{2}/U_{0}=0.05$, it shows that $\langle S^{2} \rangle=0$ for all even and $\langle S^{2} \rangle=2$ for all odd MI lobes at all values of $U_{3}/U_{0}$. This helps us to obtain the boundary equation for each odd and even MI lobes individually and they have been plotted with $U_{3}/U_{0}$ in Fig.\ref{4} (dashed lines). For $U_{2}/U_{0}=0.05$, the phase diagrams obtained via PMFA and numerically computing Eq.(\ref{mf}) are in accordance with each other for all values of $U_{3}/U_{0}$. However, for $U_{2}/U_{0}=0.3$, a minor mismatch is observed at tip of the third MI lobe corresponding to $U_{3}/U_{0}=-2.0$.
\\ \indent It has to be noted that such disparity between the two approaches are due to the less sensitivity of MFA in lower dimensions and it is not an appropriate tool to handle fluctuations properly \cite{PhysRevA.70.043628}. Further if we solve the above equations [Eq.(\ref{even}) and Eq.(\ref{odd})], which are quadratic in $\mu$ indicates that the critical hopping strength, $t_{c}/U_{0}$ denoting the location of the MI-SF phase transition (by equating $\mu_{+}$ and $\mu_{-}$) is now a function of $W$ and $U_{3}$ and increases with the strength of the three body interaction potentials. 
\\\indent The phase diagrams in Fig.\ref{4} show that the attractive three body interaction diminishes the odd-even asymmetry around the second and the fourth MI lobes, which typically 
raises one concern, namely, whether there will be any such asymmetry left beyond the fourth MI lobe. To answer that, we have displayed the occupation density for values larger than $\rho=4$ and found that the chemical potential widths for the fifth as well as the sixth MI lobes remain unaltered for different values of $U_{3}/U_{0}$ [Fig.\ref{7}(a)]. This certainly signifies the importance of such interaction potentials solely around the third MI lobe and also indicates that the asymmetry being intact beyond the fourth MI lobe. Also, for $U_{2}/U_{0}=0.05$, the transition from the spin singlet MI to the SF phase still maintains a first order character for the even MI lobes, while it shows a second order transition for the odd MI lobes at $U_{3}=-0.5$ [Fig.\ref{7}(d)]. But the MI-SF phase transition for the third ($\rho=3$) MI lobe shows a first order transition at both values of $U_{3}$, that is, $U_{3}=-1.1$ and $-1.4$ [Fig.\ref{7}(c)].
\\ \indent The mean field phase diagrams corresponding to
the ferromagnetic case are identical with that of the scalar Bose gas as obtained earlier
in Refs.\cite{PhysRevA.78.043603,PhysRevA.85.065601,PhysRevA.88.063625} for the repulsive and in Ref.\cite{PhysRevLett.109.135302} for the attractive three body interactions. All the MI lobes stabilize with increasing repulsive three body strength, $W/U_{0}$, while in the other case, only the third MI lobe stabilizes as $U_{3}/U_{0}$ is increased. We have also performed similar perturbation calculations and they are in complete agreement with the MFA phase diagrams.
%%%%%%%%%%%%%%%%%%%%%%%%%%%%%%%%%%%%%%%%%%%%
\section{Conclusion}
In this work, we have explored the consequences of both the repulsive as well as the attractive three body interaction on the phase diagrams of a spin-1 Bose Hubbard model (BHM) using mean field approximations. 
\\\indent For antiferromagnetic (AF) spin dependent interaction, we first consider both the repulsive two and three body interactions together and have found the existence of the odd-even asymmetry in the MI lobes which is in contradiction with the results obtained in Ref.\cite{arXiv:1707.08195}. However, we have noticed that the spin nematic (singlet) formation only possible at the first (second) MI lobe which were subsequently confirmed by computing the total spin eigen value and spin nematic order parameter. Further, we have confirmed that the superfluid (SF) phase is completely in transverse polar (TP) state and there is no signature of any phase transition therein which is again in contradiction with Ref.\cite{arXiv:1707.08195}. 
\\\indent But in case of only three body repulsive interaction, the phase diagram shows that the higher order MI lobes, excluding the first one, stabilize against the SF phase and there is no indication of the odd-even asymmetry in the MI lobes which are qualitatively in agreement with the results of Ref.\cite{PhysRevA.94.033623}. Thus we can conclude that the two body interaction is totally responsible for such asymmetry in the MI lobes, while the three body interaction acts as a catalyst to stabilize the insulating phases, against the SF phase.
\\\indent For an attractive three body interaction in the AF case, the third MI lobe is severely affected and it breaks odd-even asymmetry in the neighborhood of the second and fourth MI lobes at higher values of the interaction strength. Although, the behaviour of the spin eigen value and the nematic order parameter behaviour tells us that the asymmetry is still present beyond the fourth MI lobe.
\\\indent In the ferromagnetic case, the phase diagrams are similar to that of the spin-0 Bose gas for both the repulsive \cite{PhysRevA.78.043603,PhysRevA.85.065601,PhysRevA.88.063625} and the attractive \cite{PhysRevLett.109.135302} three body interactions. Finally, all these MFA phase diagrams are compared with the analytical phase diagrams obtained using perturbative expansions.    
%%%%%%%%%%%%%%%%%%%%%%%%%%%
\section*{Acknowledgments}
SB thanks DST-SERB, India for financial support under the grants no: EMR/2015/001039.
\bibliographystyle{aip}
\bibliography{referance.bib}

\begin{thebibliography}{10}

\bibitem{Greiner}
M.~Greiner, O.~Mandel, T.~Esslinger, T.~W. Hansch, and I.~Bloch,
\newblock Nature {\bf 415}, 39 (2002).

\bibitem{PhysRevLett.81.3108}
D.~Jaksch, C.~Bruder, J.~I. Cirac, C.~W. Gardiner, and P.~Zoller,
\newblock Phys. Rev. Lett. {\bf 81}, 3108 (1998).

\bibitem{PhysRevLett.80.2027}
D.~M. Stamper-Kurn, M.~R. Andrews, A.~P. Chikkatur, S.~Inouye, H.-J. Miesner,
  J.~Stenger, and W.~Ketterle,
\newblock Phys. Rev. Lett. {\bf 80}, 2027 (1998).

\bibitem{PhysRevLett.82.2228}
H.-J. Miesner, D.~M. Stamper-Kurn, J.~Stenger, S.~Inouye, A.~P. Chikkatur, and
  W.~Ketterle,
\newblock Phys. Rev. Lett. {\bf 82}, 2228 (1999).

\bibitem{PhysRevLett.81.742}
T.-L. Ho,
\newblock Phys. Rev. Lett. {\bf 81}, 742 (1998).

\bibitem{ohmi}
T.~Ohmi and K.~Machida,
\newblock J. Phys. Soc. Jpn. {\bf 67}, 1822 (1998).

\bibitem{PhysRevA.91.043620}
S.~S. Natu, J.~H. Pixley, and S.~Das~Sarma,
\newblock Phys. Rev. A {\bf 91}, 043620 (2015).

\bibitem{PhysRevB.69.094410}
M.~Snoek and F.~Zhou,
\newblock Phys. Rev. B {\bf 69}, 094410 (2004).

\bibitem{PhysRevB.77.014503}
R.~V. Pai, K.~Sheshadri, and R.~Pandit,
\newblock Phys. Rev. B {\bf 77}, 014503 (2008).

\bibitem{PhysRevA.82.063602}
L.~de~Forges~de Parny, M.~Traynard, F.~H\'ebert, V.~G. Rousseau, R.~T.
  Scalettar, and G.~G. Batrouni,
\newblock Phys. Rev. A {\bf 82}, 063602 (2010).

\bibitem{PhysRevB.84.064529}
L.~de~Forges~de Parny, F.~H\'ebert, V.~G. Rousseau, R.~T. Scalettar, and G.~G.
  Batrouni,
\newblock Phys. Rev. B {\bf 84}, 064529 (2011).

\bibitem{PhysRevB.88.104509}
L.~de~Forges~de Parny, F.~H\'ebert, V.~G. Rousseau, and G.~G. Batrouni,
\newblock Phys. Rev. B {\bf 88}, 104509 (2013).

\bibitem{PhysRevLett.102.140402}
G.~G. Batrouni, V.~G. Rousseau, and R.~T. Scalettar,
\newblock Phys. Rev. Lett. {\bf 102}, 140402 (2009).

\bibitem{PhysRevA.70.043628}
S.~Tsuchiya, S.~Kurihara, and T.~Kimura,
\newblock Phys. Rev. A {\bf 70}, 043628 (2004).

\bibitem{PhysRevLett.94.110403}
T.~Kimura, S.~Tsuchiya, and S.~Kurihara,
\newblock Phys. Rev. Lett. {\bf 94}, 110403 (2005).

\bibitem{PhysRevA.87.043624}
T.~Kimura,
\newblock Phys. Rev. A {\bf 87}, 043624 (2013).

\bibitem{PhysRevA.68.063602}
A.~Imambekov, M.~Lukin, and E.~Demler,
\newblock Phys. Rev. A {\bf 68}, 063602 (2003).

\bibitem{PhysRevLett.90.250402}
S.~K. Yip,
\newblock Phys. Rev. Lett. {\bf 90}, 250402 (2003).

\bibitem{EPL.63.505}
F.~Zhou,
\newblock EPL (Europhysics Letters) {\bf 63}, 505 (2003).

\bibitem{PhysRevLett.95.240404}
M.~Rizzi, D.~Rossini, G.~De~Chiara, S.~Montangero, and R.~Fazio,
\newblock Phys. Rev. Lett. {\bf 95}, 240404 (2005).

\bibitem{PhysRevA.83.013605}
M.~\L{}\k{a}cki, S.~Paganelli, V.~Ahufinger, A.~Sanpera, and J.~Zakrzewski,
\newblock Phys. Rev. A {\bf 83}, 013605 (2011).

\bibitem{NoorJPB}
S.~N. Nabi and S.~Basu,
\newblock J. Phys. B: At. Mol. Opt. Phys. {\bf 49}, 125301 (2016).

\bibitem{PhysRevB.95.235128}
K.~D. McAlpine, S.~Paganelli, S.~Ciuchi, A.~Sanpera, and G.~De~Chiara,
\newblock Phys. Rev. B {\bf 95}, 235128 (2017).

\bibitem{PhysRevLett.97.020401}
S.~Yi and H.~Pu,
\newblock Phys. Rev. Lett. {\bf 97}, 020401 (2006).

\bibitem{PhysRevLett.97.130404}
Y.~Kawaguchi, H.~Saito, and M.~Ueda,
\newblock Phys. Rev. Lett. {\bf 97}, 130404 (2006).

\bibitem{PhysRevB.92.054506}
C.-C. Chang, V.~G. Rousseau, R.~T. Scalettar, and G.~G. Batrouni,
\newblock Phys. Rev. B {\bf 92}, 054506 (2015).

\bibitem{NOORANDP}
S.~N. Nabi and S.~Basu,
\newblock Annalen der Physik , 1700245,
\newblock 1700245.

\bibitem{PhysRevA.93.063607}
J.~Jiang, L.~Zhao, S.-T. Wang, Z.~Chen, T.~Tang, L.-M. Duan, and Y.~Liu,
\newblock Phys. Rev. A {\bf 93}, 063607 (2016).

\bibitem{PhysRevA.94.063613}
H.~M. Hurst, J.~H. Wilson, J.~H. Pixley, I.~B. Spielman, and S.~S. Natu,
\newblock Phys. Rev. A {\bf 94}, 063613 (2016).

\bibitem{PhysRevLett.84.1066}
M.~Koashi and M.~Ueda,
\newblock Phys. Rev. Lett. {\bf 84}, 1066 (2000).

\bibitem{PhysRevLett.87.080401}
F.~Zhou,
\newblock Phys. Rev. Lett. {\bf 87}, 080401 (2001).

\bibitem{Zhang}
W.~Zhang, S.~Yi, and L.~You,
\newblock New Journal of Physics {\bf 5}, 77 (2003).

\bibitem{PhysRevB.70.184434}
F.~Zhou, M.~Snoek, J.~Wiemer, and I.~Affleck,
\newblock Phys. Rev. B {\bf 70}, 184434 (2004).

\bibitem{NoorEPL}
S.~N. Nabi and S.~Basu,
\newblock EPL (Europhysics Letters) {\bf 116}, 46001 (2016).

\bibitem{PhysRevLett.112.043001}
A.~Celi, P.~Massignan, J.~Ruseckas, N.~Goldman, I.~B. Spielman,
  G.~Juzeli\ifmmode~\bar{u}\else \={u}\fi{}nas, and M.~Lewenstein,
\newblock Phys. Rev. Lett. {\bf 112}, 043001 (2014).

\bibitem{PhysRevA.91.023608}
S.~S. Natu, X.~Li, and W.~S. Cole,
\newblock Phys. Rev. A {\bf 91}, 023608 (2015).

\bibitem{PhysRevA.93.013629}
L.~Chen, H.~Pu, and Y.~Zhang,
\newblock Phys. Rev. A {\bf 93}, 013629 (2016).

\bibitem{PhysRevA.93.023615}
K.~Sun, C.~Qu, Y.~Xu, Y.~Zhang, and C.~Zhang,
\newblock Phys. Rev. A {\bf 93}, 023615 (2016).

\bibitem{PhysRevB.93.081101}
J.~H. Pixley, S.~S. Natu, I.~B. Spielman, and S.~Das~Sarma,
\newblock Phys. Rev. B {\bf 93}, 081101 (2016).

\bibitem{PhysRevLett.117.125301}
G.~I. Martone, F.~V. Pepe, P.~Facchi, S.~Pascazio, and S.~Stringari,
\newblock Phys. Rev. Lett. {\bf 117}, 125301 (2016).

\bibitem{Will}
S.~Will, T.~Best, U.~Schneider, L.~Hackermuller, D.-S. Luhmann, and I.~Bloch,
\newblock Nature {\bf 465}, 197 (2010).

\bibitem{PhysRevA.78.043603}
B.-l. Chen, X.-b. Huang, S.-p. Kou, and Y.~Zhang,
\newblock Phys. Rev. A {\bf 78}, 043603 (2008).

\bibitem{PhysRevA.84.065601}
J.~Silva-Valencia and A.~M.~C. Souza,
\newblock Phys. Rev. A {\bf 84}, 065601 (2011).

\bibitem{PhysRevA.85.065601}
T.~Sowi\ifmmode~\acute{n}\else \'{n}\fi{}ski,
\newblock Phys. Rev. A {\bf 85}, 065601 (2012).

\bibitem{PhysRevA.88.063625}
S.~Ejima, F.~Lange, H.~Fehske, F.~Gebhard, and K.~z. M\"unster,
\newblock Phys. Rev. A {\bf 88}, 063625 (2013).

\bibitem{PhysRevA.88.023602}
K.~W. Mahmud and E.~Tiesinga,
\newblock Phys. Rev. A {\bf 88}, 023602 (2013).

\bibitem{PhysRevA.94.033623}
A.~F. Hincapie-F, R.~Franco, and J.~Silva-Valencia,
\newblock Phys. Rev. A {\bf 94}, 033623 (2016).

\bibitem{arXiv:1707.08195}
A.~F. Hincapie-F, R.~Franco, and J.~Silva-Valencia,
\newblock arXiv:1707.08195 .

\bibitem{PhysRevLett.109.135302}
A.~Safavi-Naini, J.~von Stecher, B.~Capogrosso-Sansone, and S.~T. Rittenhouse,
\newblock Phys. Rev. Lett. {\bf 109}, 135302 (2012).

\bibitem{Pethick}
C.~J. Pethick and H.~Smith,
\newblock {\em Bose-Einstein Condensation in Dilute Gases},
\newblock Cambridge University Press, Cambridge, England, 2002.

\bibitem{PhysRevLett.88.163001}
E.~Demler and F.~Zhou,
\newblock Phys. Rev. Lett. {\bf 88}, 163001 (2002).

\end{thebibliography}
\end{document}